\documentclass[%
reprint,
amsmath,amssymb,
aip, 
prstab,
]{revtex4-1}

\usepackage{graphicx}
\usepackage{dcolumn}
\usepackage{bm}

\usepackage[squaren]{SIunits}
\usepackage[dvipsnames]{xcolor}
\usepackage{amsmath}
\usepackage{nicefrac}
\AtBeginDocument{%
  \hypersetup{
    citecolor=blue,
    linkcolor=blue,   
    urlcolor=blue}}
\usepackage[dvipsnames]{xcolor}
\usepackage[normalem]{ulem}

\usepackage{natbib}
\usepackage{float}
\usepackage{textcomp}
\usepackage[colorlinks]{hyperref}
\usepackage{caption}
\usepackage{subcaption}
\DeclareCaptionFont{mysize}{\fontsize{9}{9.6}\selectfont}
\usepackage{multirow}
\usepackage{booktabs}

\begin{document}

\title{Dephasing of ion beams as Magnetic Vortex Acceleration regime transitions into a bubble-like field structure}
\author{Sahel Hakimi}
\author{Stepan S. Bulanov}
\email{sbulanov@lbl.gov}
\author{Axel Huebl}
\author{Lieselotte Obst-Huebl}
\author{Kei Nakamura}
\author{Anthony Gonsalves}
\author{Thomas Schenkel}
\author{Jeroen van Tilborg}
\author{Jean-Luc Vay}
\author{Carl B. Schroeder}
\author{Eric Esarey}
\author{Cameron R. Geddes}
\affiliation{%
 $^1$ Lawrence Berkeley National Laboratory, Berkeley, California 94720, USA\\
}%

\begin{abstract}
The interaction of an ultra-intense laser pulse with a near critical density target results in the formation of a plasma channel, a strong azimuthal magnetic field and moving vortices. An application of this is the generation of energetic and collimated ion beams via Magnetic Vortex Acceleration. The optimized regime of Magnetic Vortex Acceleration is becoming experimentally accessible with new high intensity laser beamlines coming online and advances made in near critical density target fabrication. The robustness of the acceleration mechanism with realistic experimental conditions is examined with three-dimensional simulations. Of particular interest is the acceleration performance with different laser temporal contrast conditions, in some cases leading to pre-expanded target profiles prior to the arrival of the main pulse. Preplasma effects on the structure of the accelerating fields is explored, including a detailed analysis of the ion beam properties and the efficiency of the process. Improved scaling laws for the MVA mechanism, including the laser focal spot size effects, are presented. 

\end{abstract}

\maketitle
\section{\label{sec:Intro}Introduction}
The interaction of an ultra-intense laser pulse with a near critical density (NCD) target results in the formation of a plasma channel, a strong azimuthal magnetic field and moving magnetic vortices with applications in non-linear plasma physics and astrophysics~\cite{mourou.rmp.2006,daido.rpp.2012,macchi.rmp.2013,bulanov.pu.2014,mp3report.2022.arxiv}. An application of this interaction is the generation of energetic and collimated ion beams via Magnetic Vortex Acceleration (MVA)~\cite{kuznetsov2001efficiency, bulanov2005ion, bulanov2007comment,willingale.prl.2006,fukuda.prl.2009,willingale.pop.2011}, a promising advanced laser-driven ion acceleration scheme~\cite{bulanov2016radiation}. With the recent developments in laser technology and construction of PW-class laser facilities~\cite{danson2019petawatt, gonoskov.rmp.2022} housing ultra-high intensity beamlines, such as the iP2 beamline~\cite{hakimi.pop.2022,LOH.SPIE2023} at the Berkeley Lab Laser Accelerator (BELLA) Center, experimental investigation of the MVA regime is under way~\cite{hakimi.2024}. 

In addition to the stringent laser parameter requirements, optimized target designs are required for experimental studies of MVA and verification of its scaling laws. MVA utilizes an NCD target with a thickness greater than the laser pulse length. Suitable targets may be realized in the form of supersonic gas-jets~\cite{sylla2012development, henares2019development} or liquid-jets~\cite{rehwald.natcomm.2023}, shock waves in a gas jet \cite{helle.prl.2016}, thin liquid sheets~\cite{koralek2018generation}, nano-clusters~\cite{fukuda.prl.2009}, and carbon-hydrogen (CH) foams and aerogels~\cite{prencipe2016development, manuel2021study}. The first few types of targets have the advantages of being regenerative for high repetition rate operation, and produce no debris in the experimental chamber. The latter can be fabricated with a range of densities and can be finely structured for a more controlled laser-plasma interaction, and although it is not regenerative, it can be fabricated in arrays for rastering at high repetition rates. 

Production of structured targets with foams and aerogels has been on the forefront of research and development for high energy density physics experiments and has advanced considerably in the last decades~\cite{prencipe2017targets}. Layered targets with NCD foams attached to nm scale diamond-like carbon foils have been used to increase the proton cutoff energy by shaping the laser pulse via simultaneous self-focusing and pulse front steepening \cite{bin.prl.2015}, and have been studied in simulations with the goal to enhance the acceleration by mitigating different limiting factors \cite{bulanov.prl.2015} in the NCD layer. Micron scale foils covered with microtubes have shown to increase the conversion efficiency of laser energy to accelerated ion energy \cite{passoni.prab.2016, bailly2020ion}. Microchannel arrays were proposed to enhance gamma generation through relativistically transparent magnetic filaments \cite{rinderknecht.njp.2021}. A waveguide structure with a pre-filled central channel and high density walls was studied numerically and found capable of accelerating a mono-energetic ion bunch~\cite{gong.prr.2022}. Similar waveguide structures have been studied for multi-MeV photon generation \cite{stark.prl.2016}. It is clear that precise shaping and structuring of the targets can provide a high level of control over the dynamics of the interaction given the target maintains its integrity for the main pulse interaction. The target profile can also be tailored by manipulating the temporal laser pulse contrast as was shown in Ref. \cite{rehwald.natcomm.2023}.

Maintaining target integrity is one of the main challenges of laser-matter interactions at high laser intensities. It is well known that some high power laser systems can generate pulses with less than 100~fs duration with a temporal profile that includes a nanosecond amplified spontaneous emission pedestal, a rising ramp before the main pulse on the order of 10's of picoseconds, and short prepulses with pulse lengths on the order of the main pulse length. The prepulses and pedestal will heat, ionize and expand the target, altering its density profile prior to the main interaction. In other cases the expansion of the target can lead to higher absorption of laser energy by the target electrons and ions, thus, leading to a more efficient acceleration. In all cases, the expansion rate and preplasma scale length at the time of the main pulse interaction is critical to how the dynamic will evolve subsequently. The propagation of possible shock waves in the target, motion of fast and slow electrons and thermal conduction can result in a modified target profile at the rear surface as well. This is especially important for the MVA regime as the onset of the acceleration stage is linked to the laser pulse exiting the target rear. Therefore, a density gradient at the target rear modifies the acceleration process.

In the MVA regime~\cite{bulanov2010generation, nakamura2010high, bulanov.prab.2015, sharma.scirep.2018, park2019ion, psikal.ppcf.2021, hakimi.pop.2022, garten.prr.2024}, an intense laser pulse makes a channel first in the electron density, then in the ion density by the ponderomotive pressure. As the laser propagates in this self-generated channel, it drives a strong electron current in its wake. This electron current flows through the ion background, which forces it to pinch towards the axis due to the plasma lens effect. The electron current propagation is accompanied by two effects. First, the generation of a strong magnetic field, which is intensified as the current pinches. Second, the ions are attracted by the moving electrons and form their own filament inside the electron filament. As the laser and the filaments leave the target rear, the magnetic field of the current starts to expand in the transverse direction. While doing so, it pushes the electrons at the back of the target back and sideways, displacing them from the ion core. This charge separation leads to the generation of strong, longitudinally accelerating and transversely focusing electric fields for ions. These fields accelerate and focus the ions from the filament, producing a high energy, well collimated ion beam~\cite{bulanov2010generation, bulanov.prab.2015, park2019ion, hakimi.pop.2022}. It follows from this description that the laser pulse should have enough energy to penetrate the NCD target, determined by the depletion rate of the laser energy in such a plasma. Thus, the target thickness should be equal or less than the depletion length\cite{park2019ion}, $L_{ch}$: 
\begin{equation}
    \frac{L_{ch}}{L_p}=K^{2/3}\left(\frac{2P}{P_c}\right)^{1/3}\left(\frac{n_{cr}}{n_e}\right)^{2/3},
\end{equation}
where $n_e$ is the plasma density, $P_\mathrm{c}=2 m_\mathrm{e}^2 c^5/e^2=17$ GW is the characteristic power for relativistic self-focusing, $n_\mathrm{cr}= m_\mathrm{e}\omega^2/(4\pi e^2)$ is the plasma critical density, $\lambda$, $L_p=c\tau$, $\omega$, and $P$ are the laser wavelength, pulse length, frequency, and power respectively, $e$ and $m_e$ are electron charge and mass, and $K=1/13.5$ is the geometric factor.

Previous analytical and simulation research \cite{bulanov2010generation, park2019ion} indicated that the peak ion energy in MVA scales as $P^{2/3}$,  enabling multi-100 MeV protons to be produced by PW-class lasers. In order to put these energies in the context of experimental results on ion acceleration, we note that for many years, the maximum ion energy was limited by a 100 MeV threshold and only recently it has reached 150 MeV level~\cite{ziegler2024laser}. These experimental studies used different targets and employed different laser facilities, thus operating in different ion acceleration regimes~\cite{bulanov2016radiation}, ranging from target normal sheath acceleration~\cite{wilks.pop.2001,wagner2016maximum} to shock wave acceleration~\cite{haberberger.NatPhys.2011} and RPA~\cite{kim2016radiation}, or even some combination of these regimes~\cite{higginson2018near}. The maximum ion energy produced by different acceleration mechanisms scales differently with laser power. Thus, the scaling serves as an identifying feature of the acceleration mechanisms as well as an indication which application might benefit most from it.

The MVA scaling studies relied on the assumption that the NCD target has a sharp plasma-vacuum interface at the target rear. In reality, for MVA experiments, the target, whether it be a gasjet, liquid jet or CH foam, can have an expanded profile at the front and, more importantly, at the rear surface. It is usually stated that MVA is more robust towards the laser contrast than, e.g., RPA, but the magnetic field expansion at the target rear can be significantly different for different plasma density gradient profiles at the rear surface \cite{kuznetsov2001efficiency,bulanov2010generation,nakamura2010high}. The evolution of the magnetic field as it expands in the vacuum is critically linked to the consequent formation of MVA accelerating and focusing electric fields. As it was mentioned above, the prepulse of the laser would result in preplasma formation both at the front and the rear side of the target, affecting the evolution of electric and magnetic fields. This can result in a less efficient ion acceleration.

In this paper, we study the effects of different preplasma scale lengths on the MVA mechanism. In three-dimensional particle-in-cell (3D PIC) computer simulations, we show that increasing the scale length leads to a dephasing between the protons and the MVA longitudinal accelerating electric field, ultimately resulting in significantly reduced proton peak energy. We also analyze the properties of the accelerated proton beam, i.e., maximum energy, total charge, and angular distribution. We observe that the normalized emittance of the generated proton beams is extremely low, in the 10's of nm range, due to small angular divergences and source sizes. 

The results presented in this paper indicate the importance of target design and laser contrast for efficient MVA operation. In the case when they can be optimized, the MVA is expected to produce high quality proton beams, which can potentially be used for further acceleration in different energy boosting schemes with finite energy and angle acceptance~\cite{garten.prr.2024}. 

The paper is organized as follows: In section~\ref{sec:simSetup}, 3D PIC simulation parameters and setup are presented. In section~\ref{sec:dephasing}, the dephasing of ions with respect to the longitudinal accelerating electric field in the case of a density gradient at the target rear is explored. In section~\ref{sec:beam-characterization}, we characterize the accelerated proton beams, where the transverse beam profile, angular distribution, and spectra are discussed. In section~\ref{sec:emittance}, the dependence of the proton beam emittance on the preplasma scale length is discussed. Scaling calculations are discussed in Section~\ref{sec:theory-modification} followed by concluding remarks in section~\ref{sec:conclusions}. 

\begin{figure*}
\centering
\includegraphics[width=0.95\textwidth]{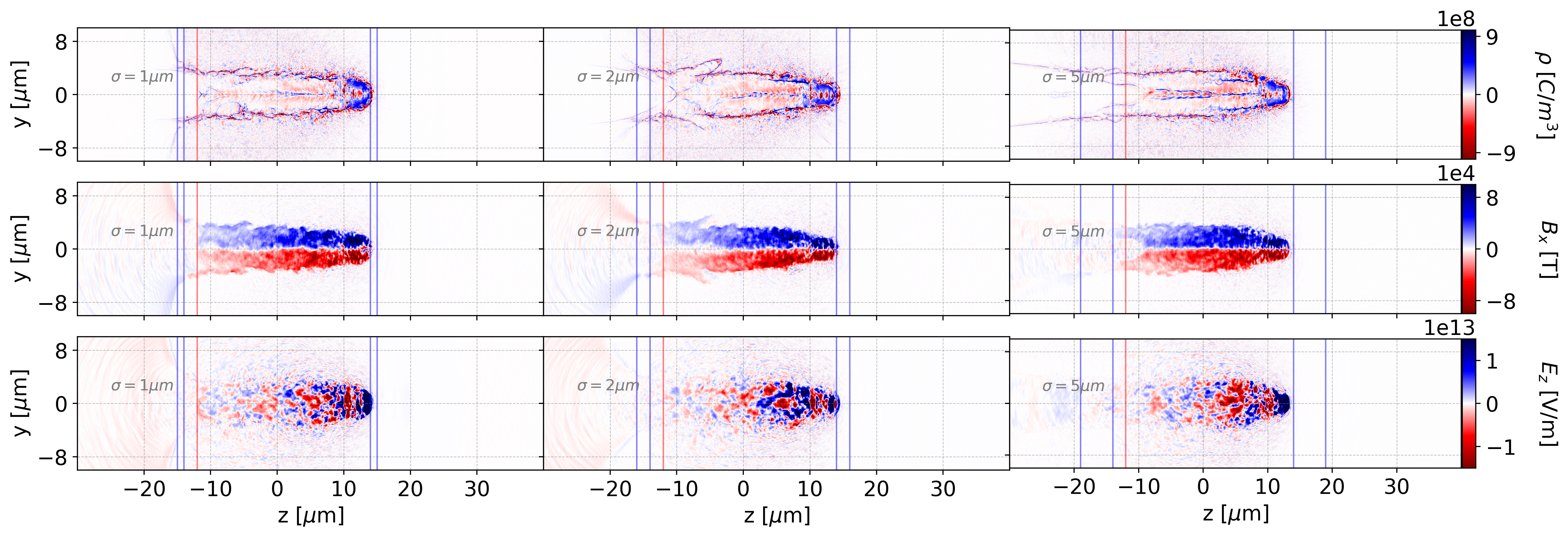}
\caption{\label{Fig:pre-plasma_Comparison_5000} 3D simulation results from the cases with a normalized laser intensity of \(a_0 = 42\). The total charge distribution (top row), azimuthal magnetic field (middle row), and longitudinal electric field (bottom row) in the (y,z) plane are shown for the cases with 1\,\(\upmu\)m (left column), 2\,\(\upmu\)m (middle column), and 5\,\(\upmu\)m (right column) at \(t=86\)\,fs, before the laser pulse exits the target. The target bulk is within \(-14<z[\upmu\)m\(]<14\) and the 1\,\(\sigma\) location of the exponential density ramps at the target front and rear sides are marked with vertical blue regions. The red vertical line marks the location of laser focus at \(z=-12\:\upmu\)m, 2\,\(\upmu\)m into the target bulk from the front surface, for all cases. A density channel and a strong azimuthal magnetic field, representative of the MVA mechanism, is observed in all cases.}
\end{figure*}

\begin{figure*}
\centering
\includegraphics[width=0.95\textwidth]{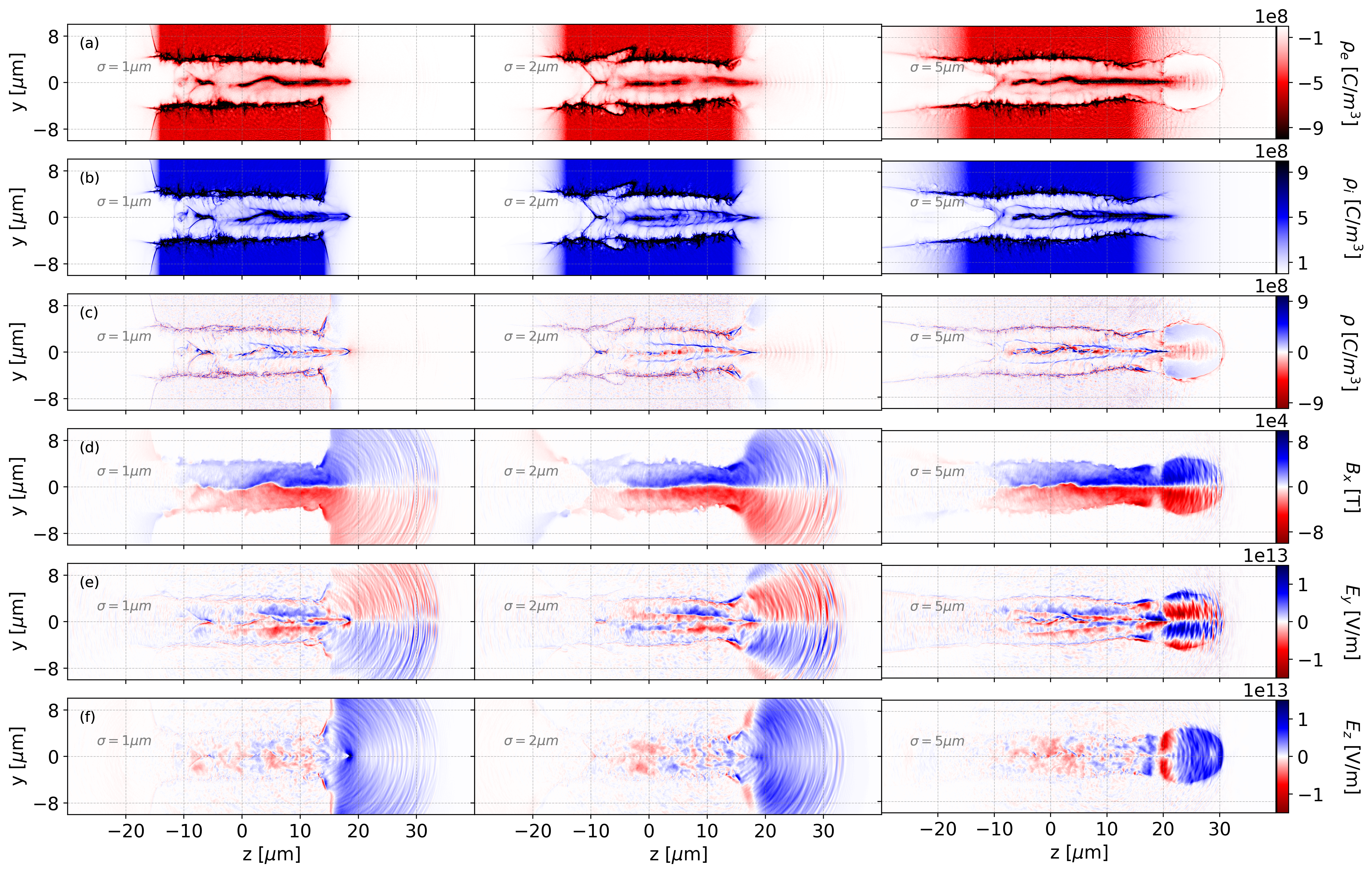}
\caption{\label{Fig:pre-plasma_Comparison_6500} The electron density distribution (a), proton distribution (b), total charge distribution (c), azimuthal magnetic field (d), transverse (e) and longitudinal (f) electric fields in the $(y,z)$ plane are shown for the cases with 1\,\(\upmu\)m (left column), 2\,\(\upmu\)m (middle column), and 5\,\(\upmu\)m (right column) at \(t=154\)\,fs, after the laser pulse exits the target. The effects of the dense preplasma region at the target rear is evident in the density and field distributions of the 5\,\(\upmu\)m case. Here, a bubble-like structure is formed from the background electron plasma that contains the fields; hence, giving rise to stronger longitudinal and transverse fields in the region spanning \(22<z[\upmu m]<30\) along the $z$-axis. However, as electrons move backward and accumulate around \(z = 20\,\upmu\)m, a thin region of decelerating longitudinal electric field for ions is formed, overlapping with the head of the ion filament. 
} 
\end{figure*}

\begin{figure}[htbp]
\begin{center}
\includegraphics[width=0.9\columnwidth]{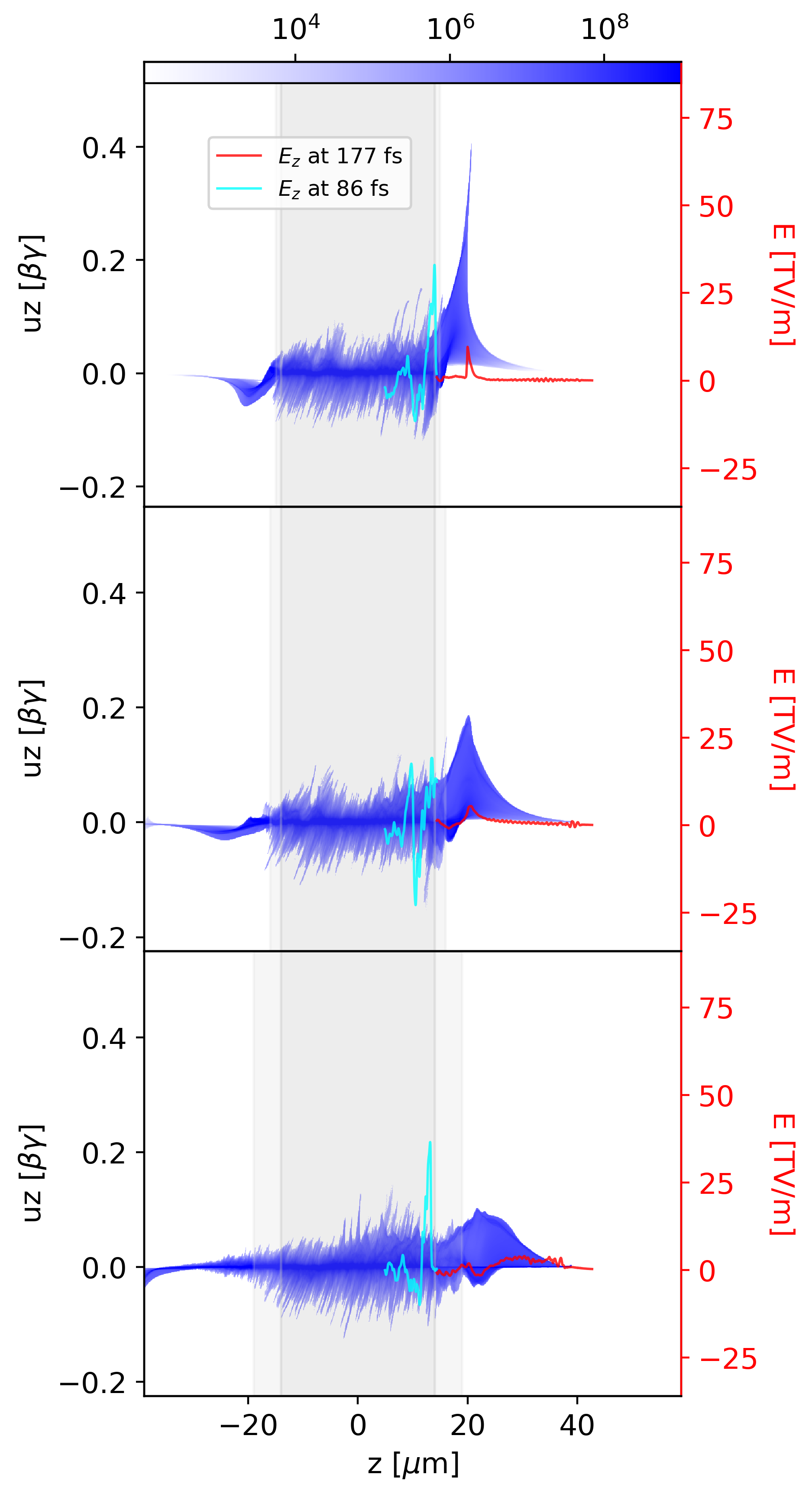}
\caption{The phase space ($z,u_z$) of protons at $t=177$\,fs in three cases of $\sigma$=1 (top), 2 (middle), and 5 (bottom) $\upmu$m, and the lineouts of the longitudinal electric field along the laser propagation axis, averaged over $\approx 2 \upmu$m in the transverse direction, at $t=86$\,fs (light blue), before the laser pulse exits the target rear, and at $t=177$\,fs (red) are shown.}
\label{Fig:phasespace}
\end{center}
\end{figure}

\section{\label{sec:simSetup} 3D Simulation parameters }

In order to study the dependence of the MVA mechanism on the preplasma scale length and to characterize the produced proton beams, a series of 3D PIC simulations were performed using the WarpX code~\cite{fedeli.proc.2022, warpx.zenodo.2018}. The parameters of the laser pulse were chosen to represent, as close as possible, what is experimentally achievable at the BELLA iP2 beamline~\cite{nakamura.ieee.2017, hakimi.pop.2022, LOH.SPIE2023}. This means we limited our consideration to two laser configurations: one with a double plasma mirror (DPM) setup featuring 24~J of laser energy on target and one without DPM featuring 40~J of laser energy on target. The target parameters were also optimized for the BELLA iP2 beamline. The NCD target is assumed as a fully ionized uniform plasma in the target bulk with symmetric exponential density gradients (preplasma), $n_{pp}=n_0 \exp[-(z-z_b)/\sigma]$, at both the front and rear sides of the target, that were cut off to zero 25\,$\upmu$m from the surfaces. 

The dimensions of the simulation box are -12\,$\upmu$m to 12\,$\upmu$m in both transverse directions ($x$ and $y$) and -40\,$\upmu$m to 75\,$\upmu$m in the longitudinal direction ($z$) with 1024, 1024, and 4864 cells in $x$, $y$ and $z$ respectively, providing a resolution of \(\lambda_\mathrm{p} /24\) in all three axes.  The laser is linearly polarized in x with the virtual antenna positioned at -39\,$\upmu$m, 1\,$\upmu$m from the left boundary of the simulation window. The laser pulse has both a transverse and longitudinal Gaussian profile and propagates along the $z$-axis. The laser wavelength is 815\,nm and the pulse duration is 35\,fs. The laser pulse is tightly focused 2\,$\upmu$m behind the target front surface at $z = -12\,\upmu$m to a FWHM of $2.5\,\upmu$m.

The peak laser intensity is \(6.05 \times  10^{21} \mathrm{W/cm^2}\) corresponding to \(a_0 = 55 ~\mbox{and}~ E_l = 40\)\,J without contrast cleaning and \(3.63 \times  10^{21} \mathrm{W/cm}^2\) corresponding to \(a_0 = 42 ~\mbox{and}~ E_l=24\)~J with contrast cleaning using the DPM setup. The target was designed with \(n_\mathrm{e} = 2 n_\mathrm{cr}\) density and \(L_\mathrm{ch} = 28\,\upmu\)m thickness to ensure that the laser pulse is able to propagate through it for all the considered intensities and preplasma scale lengths. The preplasma scale length was assumed to be $\sigma=$1, 2 and 5~$\upmu$m. Simulations were performed with two types of targets: pre-ionized pure hydrogen and Liquid Crystal Target 8CB (4-octyl-4’-cyanobiphenyl) \(C_{21}H_{25}N\) (LCT) modeled as \(C_{22}H_{25}\) for simplicity.  Other multi-species targets that provide a similar density and mixing ratio can be considered as well. Simulations were performed with \(\theta=0^\circ\) and $25^\circ$ angle of incidence. In the following sections, unless otherwise noted, results are shown for the simulations with \(a_0 = 42\) at \(0^\circ\) angle of incidence onto pure hydrogen target for three cases of preplasma scale lengths \(\sigma = 1, 2, 5\)~$\upmu$m. However, the main features discussed are observed in other simulations as well.  

\begin{figure*}
\centering
\includegraphics[width=1\textwidth]{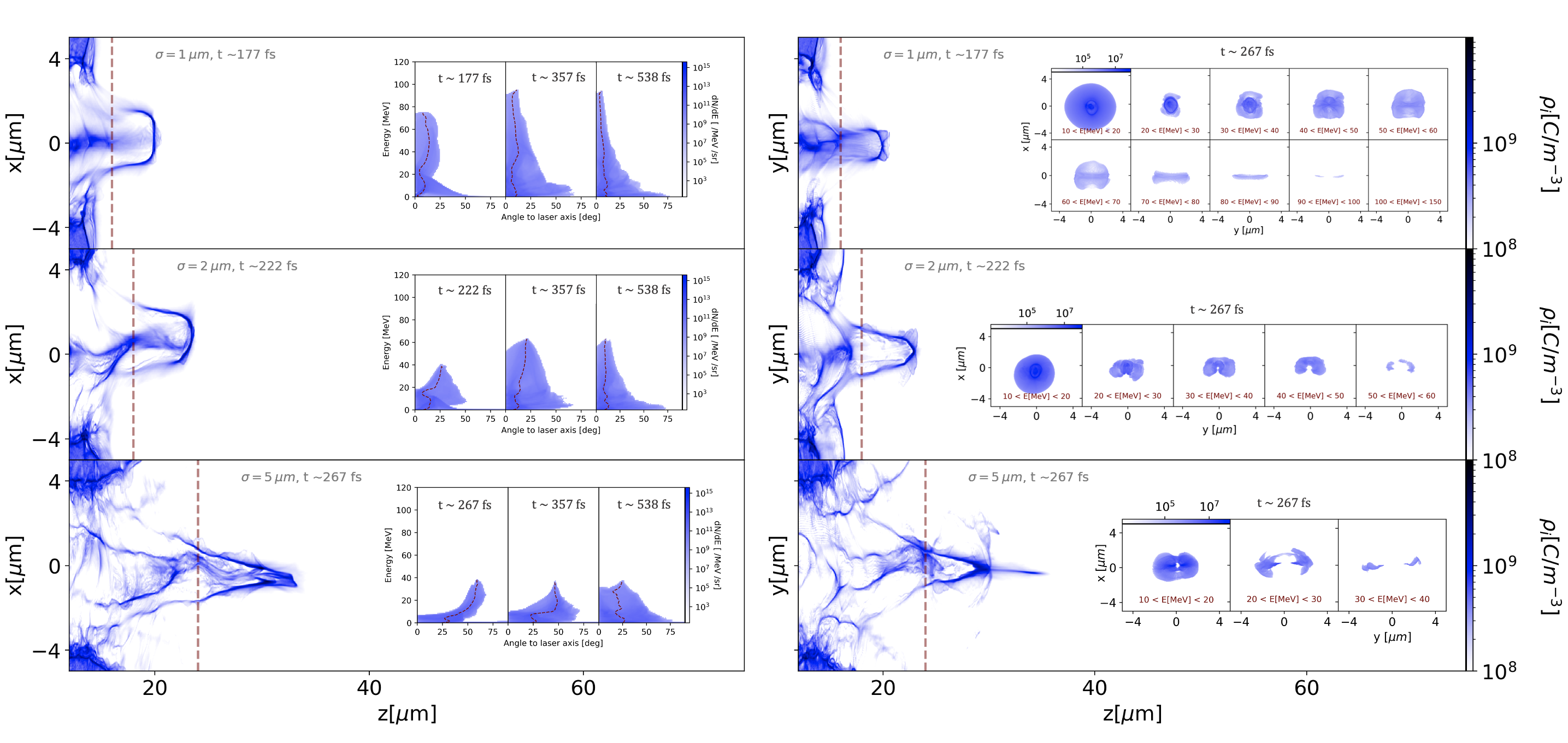}
\caption{\label{Fig:outgoingBeams}
The accelerated proton beams are shown in both (x,z) and (y,z) planes. As the preplasma scale length is increased, the laser pulse travels through an effectively thicker target and the accelerated stage is slightly delayed. Hence, the accelerated proton beams are shown at $t=177$, 222, and 267~fs for the three cases of $\sigma$~=~1,2, and 5~$\upmu$m respectively. The vertical red dashed line marks the location of \(2\sigma\) at \(z=16\), 18 and 24~$\upmu$m. The inset plots in the left columns show the angularly resolved energy spectra of the selected portion of the accelerated proton beams at \(-5>(x,y)[\upmu m]>5\) and \(z[\upmu m]>2\sigma\) at several times. The inset plots in the right columns show the integrated spatial distribution of accelerated protons within the specified energy ranges in the (x,y) plane. For each case, it is observed that the higher energy portion of the beam is spread more in the $y$ direction. }
\end{figure*}

\section{\label{sec:dephasing} Dephasing of ions in case of a density gradient profile at the rear side} 

For the simulation with \(a_0 = 42\) at \(0^\circ\) angle of incidence onto pure hydrogen target, the distributions of the charge density, the $x$-component (transverse) of the magnetic field, and the $z$-component (longitudinal) of the electric field in the $(y,z)$ plane are shown in Fig. \ref{Fig:pre-plasma_Comparison_5000} with the three different preplasma scale lengths at \(t=86\)~fs, before the laser pulse exits the target rear. Here, the referenced times are given relative to the laser focus at 2\,$\upmu$m behind the target front surface, marked with a vertical red line. The bulk of the target is at \(-14>z[\upmu\)m\(]>14\) and the exponential density ramps, up to 1\,$\sigma$, are marked with two blue vertical lines at the front and rear surfaces for each case. The formation of the density channel and the central filament, consisting of electrons and ions, the transverse magnetic field, and the longitudinal electric field are typical for the MVA mechanism and relatively similar in all cases.

Shortly after the laser exits the target rear, at \(t=154\)~fs, the differences in the density and field distributions start to become apparent as shown in Fig.~\ref{Fig:pre-plasma_Comparison_6500}. While the $\sigma$~= 1 and 2\,$\upmu$m cases look fairly similar, the longer density ramp of the $\sigma=5\,\upmu$m case leads to different density distributions and field structures at the target rear after the laser has propagated through the target. Here, a cloud of electrons is formed from the considerable background plasma distribution, resembling a ``blowout bubble", a phenomenon known from underdense plasmas in wakefield acceleration~\cite{PukhovMtV2002,esarey2009physics}. As the laser pulse propagates in this region, electrons are pushed out and away due to the laser's ponderomotive force. The dense preplasma does not allow the fields to expand in the transverse direction, maintaining the bubble-like structure, where the plasma density in the walls is high enough to contain the electromagnetic fields inside. However, there are several notable differences from the wakefield ``blowout bubble". First, the accelerating region of the longitudinal electric field, $E_z$, for ions, is much longer than the decelerating region. This is linked to the fact that the plasma has a steep down gradient, which results in the elongation of the front portion of the ``bubble". Second, the electron filament extends further into the preplasma than the ion filament, as can be seen from  Fig.~\ref{Fig:pre-plasma_Comparison_6500}, which is especially visible in the charge distribution plots. This results in the band-like transverse electric field structure, where the field of the electron filament is superimposed on the field of the ``bubble". Thus, the typically de-focusing, for ions, transverse electric field in the ``bubble", becomes focusing near the axis, as the electron filament attraction for ions is stronger than the repulsion of the background ions. Thus, due to the unique combination of a strong electron current and the plasma density down gradient, the bubble-like field structure provides both acceleration and focusing for ions. Another consequence of the 5~$\upmu$m preplasma scale length is that the entire accelerating field is in the region \(22<z[\upmu m]<30\) along the z-axis, while the peak of the accelerating field for $\sigma$~= 1 and 2~$\upmu$m cases is around the 20~$\upmu$m mark along the z-axis. Hence, the ions from the filament are not exposed to this field, which results in much lower peak proton energies observed in simulations. Thus, dephasing of the ions and the longitudinal electric field plays a critical role in reduction of the ion energy, resulting in the $\sigma$~= 5~$\upmu$m case generating almost three times smaller cut-off energies compared to the 1~$\upmu$m case. It should be noted that the contained accelerating and focusing fields in the $\sigma$~= 5~$\upmu$m case are stronger compared to the other two cases and since they are confined due to the high density walls of the bubble-like structure, they last nearly 100~fs longer. 

\begin{figure*}[htbp]
\begin{center}
\includegraphics[width=1\textwidth]{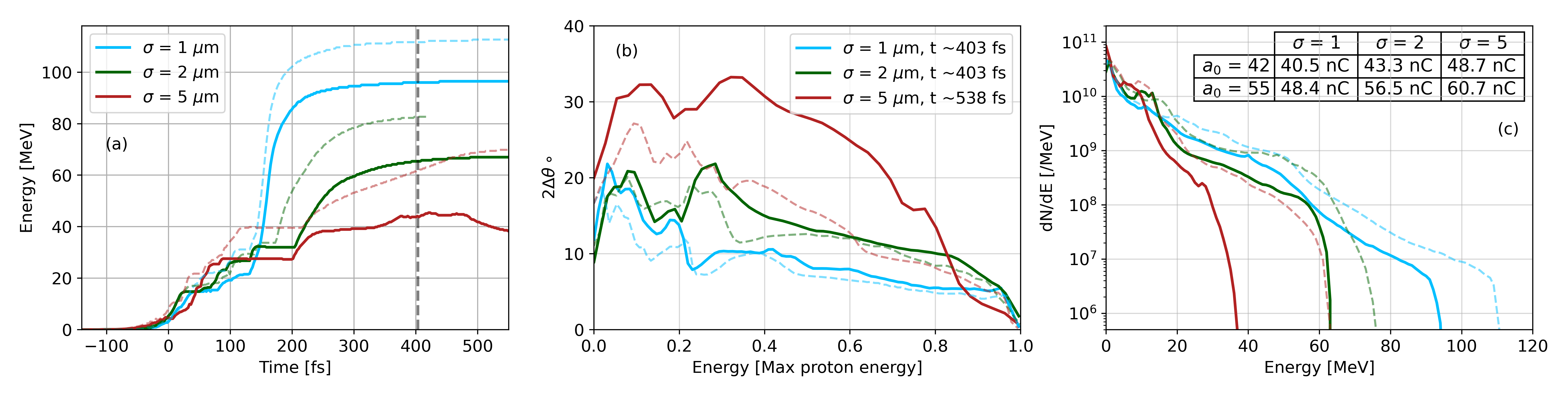}
\caption{(a) Maximum proton energy as a function of time for different preplamsa scale lengths of 1~$\mu$m (blue), 2~$\mu$m (green) and 5~$\mu$m (maroon). Simulations with \(a_0 = 42\) are shown with solid lines and simulations with \(a_0 = 55\) are shown with dashed lines. (b) Proton beam divergence as a function of normalized kinetic energy and (c) proton kinetic energy spectra are shown. The inset table shows the total charge of the proton beam at \(t=403\)~fs relative to the time, which laser focus at $z=-12$~$\upmu$m. The proton energy has converged at this time as shown in panel (a) with a vertical dashed line at \(t=403\)~fs.} 
\label{Fig:SpectraNdivergence}
\end{center}
\end{figure*}

The structure of the central filament is very similar in all three cases since the filament formation is due to the self-channeling of the intense laser pulse through the bulk of the NCD plasma. As it can be seen from Fig. \ref{Fig:pre-plasma_Comparison_5000}, this part of the interaction does not strongly depend on the preplasma scale length.

The phase space $(z,u_z)$ overlaid with $E_z$ lineouts at \(t=86\)~fs, before the laser pulse exits the target rear, and \(t=177\)~fs, for the three cases of preplasma scale length, shown in Fig.~\ref{Fig:phasespace}, further illustrates the dephasing of the proton beam with the longitudinal electric field. The $E_z$ lineouts are averaged over 100~cells, equivalent to about 2~$\upmu$m in the transverse direction. In all three cases, one can see the generation of a strong longitudinal electric field, which reaches the values of 30, 25, and 35 TV/m at $t=86$~fs for \(\sigma=1, 2, 5\)~\(\upmu\)m respectively. However, this field decays to much lower values of approximately 5 TV/m at $t=177$~fs in all cases. There is a significant difference between the 1 and 2 $\mu$m cases with the 5~$\mu$m case in terms of the longitudinal field evolution. Whereas the peaks of the former two move from $z=14$~$\mu$m to $z=22$~$\mu$m over this time interval, the later peak moves from $z=14$~$\mu$m to $z=30$~$\mu$m. This leads to a different form of the proton phase space distribution as seen in Fig.~\ref{Fig:phasespace}. First, in $\sigma$~= 1~$\mu$m case, $u_z$ reaches the value of 0.4 and this peak is just behind the peak of the longitudinal electric field. Almost the same behaviour is seen in the case of $\sigma=2$~$\mu$m, where the maximum value of $u_z$ is 0.25. The lower value of $u_z$ can be attributed to lower values of the field amplitude. In the third case, $\sigma=5$~$\mu$m, the maximum values of $u_z$ is 0.15 and the position of this peak is well behind the maximum of the field. Therefore, there is a significant dephasing of the protons and the accelerating longitudinal electric field, resulting in reduced efficiency of the MVA mechanism when a large preplasma scale is present at the target rear.   

\begin{figure*}[htbp]
\begin{center}
\includegraphics[width=1\textwidth]{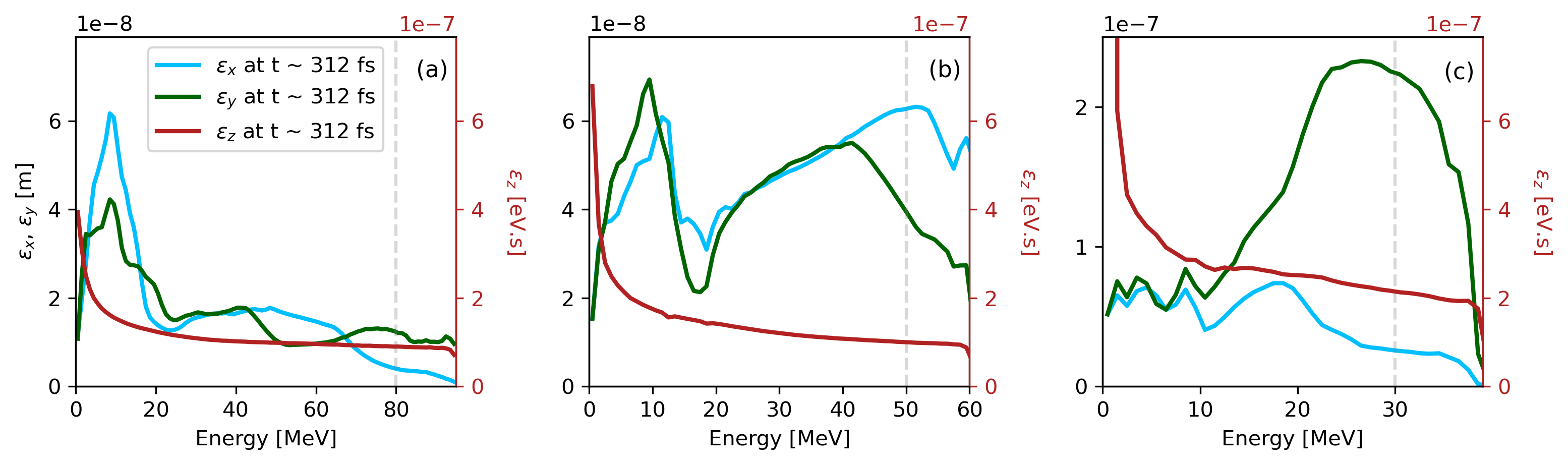}
\caption{The emittances of the accelerated proton beams for $\sigma=1,2,$ and $5$ $\mu$m. Emittance in x and y is shown on the left axis and emittance in z is shown on the right axis.}
\label{Fig:emittance}
\end{center}
\end{figure*}

\section{\label{sec:beam-characterization} Ion beam characterization }

Accelerated proton beams from the three different preplasma cases are shown in Fig.~\ref{Fig:outgoingBeams} at an early time during the acceleration stage. Since the overall target length, including the preplasma portion at the target front and rear increases as the scale length is increased, the onset of the acceleration stage increases as well, as seen in Fig.~\ref{Fig:SpectraNdivergence} (a). Hence, the accelerated ion beam properties are studied at \( t = 177, 222, 267 \)~fs for the cases of \(\sigma = 1, 2, 5 ~\upmu\)m respectively. Further analysis of the angular distribution, transverse beam profiles, beam divergence, kinetic energy spectra and emittance of the accelerated ion beams, in the inset of Fig.~\ref{Fig:outgoingBeams}, in Fig.~\ref{Fig:SpectraNdivergence} (b,c) and in Fig.~\ref{Fig:emittance}, are performed on a selection of ions within the \(-5< (x,y) [\upmu m]<5\) and \(z [\upmu m] > 16, 18, 24 \) for the cases of \(\sigma = 1, 2, 5 ~\upmu\)m respectively. Longitudinally, particles beyond the \(2\sigma\) mark are chosen to ensure beam characterization analysis is done for the accelerated ion beam and avoid selecting background ions. The red dashed line marks this z location for each case and shows the portion of the accelerated ion beams selected in each case. Here, the left and right panels show the density distribution in the (x,z) and (y,z) planes respectively to show the symmetry of the beams. In the \(\sigma = 1 ~\upmu\)m case, the ion beam has a flat front surface with two additional lobes of higher energy particles present in the (y,z) plane. The \(\sigma = 2 ~\upmu\)m case is relatively similar. In contrast, the \(\sigma = 5 ~\upmu\)m case, features a needle-like beam due to the strong transverse electric fields in this case. Although not at the highest energy, a small ion population expanding in the (y,z) plane is still present in this case. The insets in the left panels show the angularly resolved energy spectra at the same time and two future times, \( t = 357, 538 \)~fs to highlight how the proton beams evolve in time for each case. The red dashed lines show the average angle per energy bin for the selected proton beams. The integrated spatial distribution of accelerated protons in the (x,y) plane for selected energies with $\Delta E=10$~MeV are also shown in the insets of the right panels. It should be mentioned that these features are consistently seen in simulations at higher intensity, at non-zero angle of incidences, and with composite targets. 

In Fig.~\ref{Fig:SpectraNdivergence} we show the evolution of the peak proton energy (Fig.~\ref{Fig:SpectraNdivergence}a), divergence (Fig.~\ref{Fig:SpectraNdivergence}b), and the spectra of protons (Fig.~\ref{Fig:SpectraNdivergence}c) for each preplasma scale length at two values of $a_0$, as indicated in the inset in Fig.~\ref{Fig:SpectraNdivergence}c, where the total charges of accelerated proton beams are also given. The results of the simulations clearly show the process of acceleration by the longitudinal electric field at the back of the target. All the curves in Fig.~\ref{Fig:SpectraNdivergence}a demonstrate the initial acceleration of protons to around $20-30$~MeV, which can be attributed to the filament generation. These protons then propagate inside the filament until they start to experience the effects of the longitudinal electric field generated at the target rear, which takes place around $120-200$~fs depending on the preplasma scale length. Subsequently, the main bulk of the acceleration happens, accelerating protons to their maximum energy.

It can be seen that the stage where the filament forms is very similar for all the considered cases of $\sigma$ and $a_0$. The main difference stems from the interaction with the longitudinal electric field. The longer the preplasma scale length, the longer it takes the protons to reach the accelerating field and as a result, the energy gain is smaller. For the $\sigma=1$~$\mu$m case, the acceleration by the longitudinal electric field starts around 120\,fs, whereas for the $\sigma=5\,\upmu$m case, it starts around 200\,fs. The behavior of the angular distribution (Fig.~\ref{Fig:SpectraNdivergence}b) can also be explained from the dephasing. For longer preplasma scale lengths the energy gain is smaller, but the transverse momentum gain is similar. This leads to the increased angular divergence for higher values of $\sigma$. 

The proton spectra are exponentially decaying, which means the number of protons at high energies is significantly lower than that at low energies. Therefore, while the overall charge of proton beams is 40-60\,nC, the charge of the high energy parts of the beam is significantly lower. Around 1\,\% of protons go to the highest 10 MeV energy bin, resulting in 400\,pC charge.

One remarkable property of MVA generated proton beams is their small transverse size. As can be seen in Fig.~\ref{Fig:outgoingBeams}, where we zoom in on the tip of the proton filament. This is where the accelerated protons mainly originate from. The tip looks like a thin shell, couple of microns wide and less than a micron thick. The insets show the angular distribution of protons as well as the average angle, with respect to the laser axis, for given proton energies.  

In the simulations shown in Fig.~\ref{Fig:SpectraNdivergence}, we observe an energy transfer between 3.9 and 5.3~\% from the initial laser energy to the proton beam energy at the time when proton energy has reached its maximum and has converged. Most of the energy resides in the low energy part of the beam. For the simulation with \(a_0 = 42\) and \(\sigma = 1 \mu m\), the total proton beam energy is 0.94~J, with nearly 82~\% of it in $0<E_p<20$~MeV. Higher energy protons receive significantly smaller share of the total beam energy from 9.5~\% for $20<E_p<40$~MeV, 6.5~\% for $40<E_p<60$~MeV, 1.6~\% for $60<E_p<80$~MeV, to less than 0.5~\% in $E_p>80$~MeV. 

\section{\label{sec:emittance} Emittance}

It is well known that laser plasma interactions produce low emittance proton beams, as was shown in Ref.~\cite{cowan2004ultralow} for different laser-target interaction parameters, and thus, for a different acceleration mechanism. As it is mentioned above, the MVA mechanism produces proton beams in a very limited volume with a small angular spread. This, in principle, should lead to low values of the transverse and longitudinal emittances.

Normalized transverse emittances, $\epsilon_x$~[m] and $\epsilon_y$~[m], and longitudinal emittance, $\epsilon_z$~[eV.s], of the accelerated proton beams, within \(-5< (x,y) [\upmu m]<5\) and \(z [\upmu m] > 2\sigma \), for the cases of \(\sigma = 1, 2, 5 ~\upmu\)m, calculated for specific energy bins with $\Delta E=1$ MeV, at \(t=312\)~fs, are shown in Fig. \ref{Fig:emittance}. Longitudinal emittance looks very similar in all cases, while transverse emittances, specially $\epsilon_y$, increase for the longer preplasma scale length cases. However, in all cases, the values of transverse emittances are very small, in the submicron range. This probably makes MVA an interesting source option for an injection into a conventional or plasma based accelerator \cite{garten.prr.2024}. 

The evolution of the proton beam transverse and longitudinal emittances for a specific energy bin near the cutoff energy of each case, marked with dashed gray lines in Fig. \ref{Fig:emittance}, was studied. Although some change in emittance values is present as the proton beam experiences late stages of acceleration and focusing, and then starts its ballistic propagation, these changes are relatively small. Moreover, the emittance profile almost follows that of the proton angular distribution. High energy part of the distribution can be attributed to the achromatic acceleration characterized by the same divergence angle. This is also true for the emittance dependence on proton energy. 

\section{\label{sec:theory-modification} Scaling calculations }

\begin{figure*}[htbp]
\begin{center}
\includegraphics[width=1\textwidth]{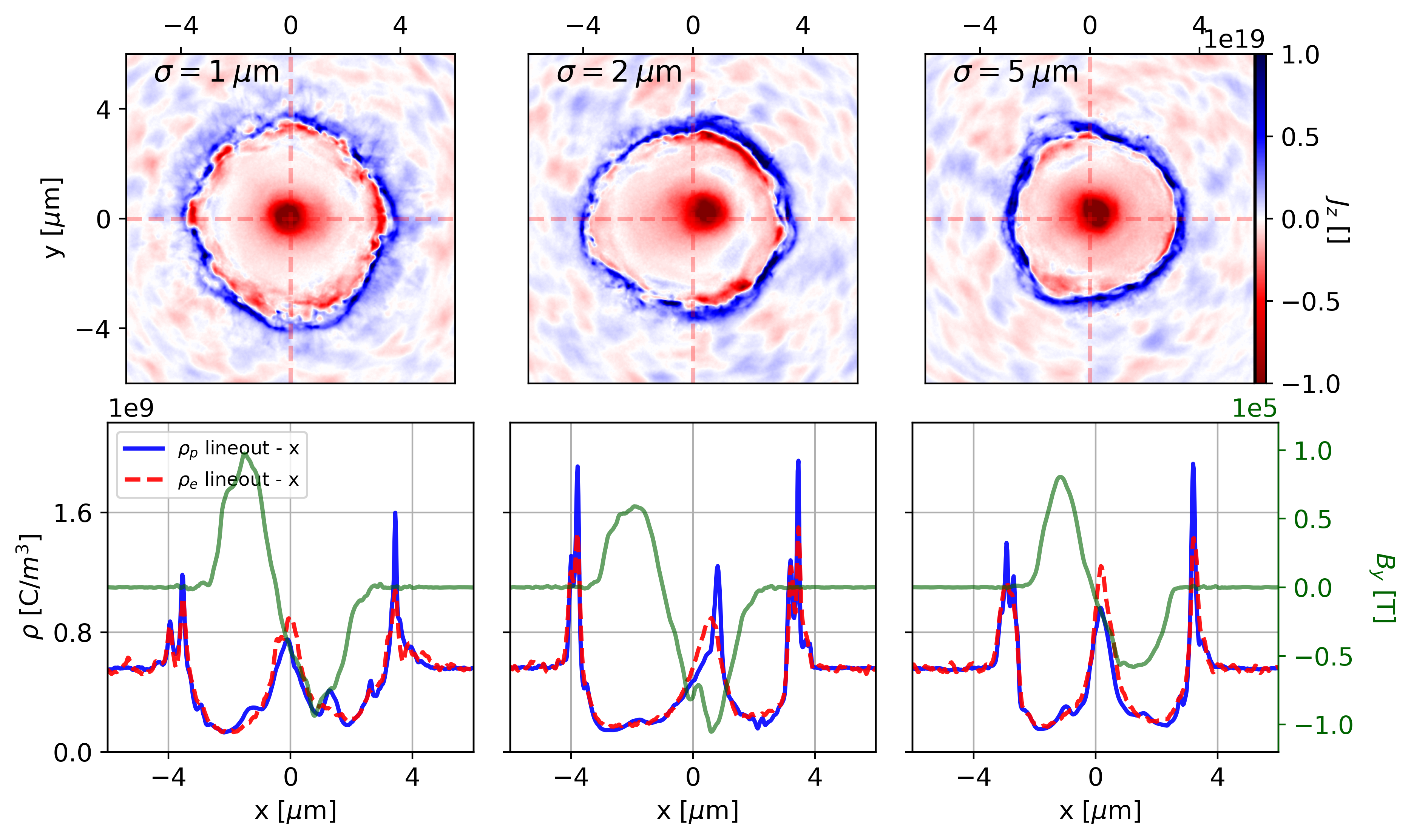}
\caption{Electric current density, $J_z$ (top), electron and proton density lineouts, $\rho_p$ (blue) and $\rho_e$ (red), (bottom, left axis) and transverse magnetic field lineouts (bottom, right axis) are shown for the three cases of preplasma scale length $\sigma$ = 1 (left), 2 (middle), 5 (right)~$\upmu$m. The transverse current density slice is integrated from $z=11$ to $z=13\upmu$m, near the target rear surface, at $t=154$~fs (same as Fig.~\ref{Fig:pre-plasma_Comparison_6500}). A strong channel is still formed in the 5~$\upmu$m case. The electron and proton channels and central filament are shown to be overlapped. The magnetic field strength inside the channel, at this time, is similar in all three cases.}
\label{Fig:current_2D}
\end{center}
\end{figure*}

In what follows we refine the analytical estimate for the magnetic field scaling with laser power. Since the magnetic field in the channel is generated by the electron current flowing in the wake of the laser pulse, the magnitude of the field is determined by the density of electrons in this current filament as well as its radius. As it was described in previous publications \cite{bulanov2010generation,bulanov.prab.2015,park2019ion}, the channel in electron density is generated by the ponderomotive push of the laser pulse experienced by plasma electrons. However, as these electrons start to move in the direction transverse to the laser propagation axis, they begin to experience the attraction by the ion column. The balance of these two effects determines the radius of the channel, which the laser pulse is able to create in the NCD plasma, $R_{ch}=(\lambda/\pi)(n_{cr}/n_e)^{1/3}(2P_{ch}/KP_c)^{1/6}$, where $P_{ch}$ is the power of the laser inside the channel. We note that $P_{ch}$ can be different from the laser power before the interaction, since not all of the laser can be coupled to the self-generated channel in general. As the walls of the channel are dense enough to prevent the laser field from leaking through, the propagation of the laser in such a self-generated channel can be described in the framework of the electromagnetic wave propagation in a waveguide \cite{bulanov2010generation}. This allows to determine the amplitude of the laser electromagnetic field in the channel, $a_{ch}=(2P_{ch}/KP_c)^{1/3} (n_e/n_{cr})^{1/3}$.  During the propagation of the laser pulse in the channel, some electrons are injected in the wake behind the pulse and accelerated in the forward direction. The results of 3D PIC simulations, reported in the previous sections, indicate that there is no plasma wave behind those electrons, which means the total number of these accelerated electrons corresponds to the beamloading limit~\cite{esarey2009physics} and can be written as:
\begin{equation}
    N_{max}=\frac{\lambda^3 n_{cr}}{4\sqrt{2}\pi^2}\left(\frac{2P_{ch}}{KP_c}\right)^{2/3}\left(\frac{n_e}{n_{cr}}\right)^{1/6}.
\end{equation}
It scales as $\sim P_{ch}^{2/3}n_e^{1/6}$, and for a 1~PW laser pulse propagating in a $2n_{cr}$ plasma, it is approximately 45~nC. Since the distribution of protons in the channel starts to follow the one of the electrons as the interaction evolves, the value of $N_{max}$ can serve as an estimate for the total number of accelerated protons. In Fig. \ref{Fig:current_2D} we show the lineouts of the electron and proton density distributions for different preplasma scale lengths, where one can see they almost coincide.  

If we use $N_{max}$ to calculate the average plasma density in the channel, where the electron current is present, we get $\overline{n_e}=n_e/4\sqrt{2}\pi$. As the electrons move through the proton background they become pinched due to the plasma lens effect and the magnetic field is intensified (see Fig. \ref{Fig:current_2D}). Following the calculations of Ref. \cite{bulanov.prab.2015}, we obtain the following result for the magnetic field generated by the pinched electron current:
\begin{equation}
    B=1.1\frac{m_e c}{\lambda}\sqrt{\frac{P_{ch}}{P_c}}.
\end{equation}
This expression yields magnetic field strength of about $\sim 0.6$ MT for a 1~PW laser pulse. Thus, the electric field generated at the back of the target, according to the MVA model, would be of the same order, i.e., $\sim 170$~TeV/m. The $\sqrt{P}$ scaling of the magnetic field amplitude is in good agreement with 3D PIC simulation results and that of Ref. \cite{park2019ion}.   

We note, though, the estimates for the magnetic field in the channel and the accelerating electric field at the back of the target are higher than the values observed in 3D PIC simulations. This is due to the fact that either the laser pulse couples non-ideally to the NCD plasma in simulations or that for the magnetic field calculation, we assume the pinched electron current has a homogeneously charged cylinder shape, whereas in simulations it has Gaussian-like transverse profiles. Taking this shape into account leads to the reduction of the field amplitude by a factor of 0.6, while the scaling with laser power stays the same.  

Let us take into account the non-ideal coupling of the laser pulse to the NCD plasma. As it was mentioned above the radius of the self-generated channel in plasma, $R_{ch}$, is determined by the plasma density and laser power. However, this laser power was assumed to be inside the channel, whereas there can be a significant power loss as the laser pulse couples to the plasma. If the focal spot at the front of the target is larger than this radius, $w_0>R_{ch}$, then the amount of laser power coupled to the NCD plasma can be estimated as $P_{ch}=(R_{ch}/w_0)^2P$. Therefore, the radius, $R_{ch}$, can be determined from the equation:
\begin{equation}
    \frac{\pi R_{ch}}{\lambda}=\left(\frac{n_{cr}}{n_e}\right)^{1/3} \left(\frac{2(R_{ch}/w_0)^2P}{K P_c}\right)^{1/6}.
\end{equation}
The solution of this equation is:
\begin{equation}
    \frac{R_{ch}}{\lambda}=\left(\frac{n_{cr}\lambda}{\pi^3 n_e w_0}\right)^{1/2}\left(\frac{2P}{K P_c}\right)^{1/4}.
\end{equation}
Thus, for $w_0>R_{ch}$, the radius of the channel depends stronger on plasma density, $\sim n_e^{-1/2}$ (instead of $\sim n_e^{-1/3}$), and on laser power, $\sim P^{1/4}$ (instead of $\sim P^{1/6}$). In addition, there is a dependence on the spot size, $\sim w_0^{-1/2}$.

The non-ideal coupling leads to the modification of optimal target thickness, which is given by the expression
\begin{equation}
    \frac{L_{ch,0}}{L_p}=\left(\frac{n_{cr}}{n_e}\right)^{2/3}\left(\frac{2P}{K P_c}\right)^{1/3}K.
\end{equation}
Taking into account $w_0>R_{ch}$, we obtain
\begin{equation}
    \frac{L_{ch}}{L_p}=\frac{K n_{cr}\lambda}{\pi n_e w_0}\left(\frac{2P}{K P_c}\right)^{1/2}.
\end{equation}
For a 1 PW laser pulse focused with $w_0=2.5$ $\mu$m and $n_e=2 n_{cr}$, $L_{ch}=0.87 L_{ch,0}$, which gives 38~$\mu$m instead of 44~$\mu$m. 

Another important consequence of non-ideal coupling is the reduction of both the magnetic field and the number of accelerated electrons. If we take into account the reduction of laser power in the channel, then
\begin{equation}
    B=2.5\frac{m_e c}{\lambda}\left(\frac{\lambda}{w_0}\right)^{3/2} \left(\frac{n_{cr}}{\pi^3 n_e}\right)^{1/2} \left(\frac{P}{P_c}\right)^{3/4}.
\end{equation}
\begin{equation}
    N_{max}=\frac{\lambda^3 n_{cr}}{4\sqrt{2}\pi^4}\frac{2P}{KP_c}\left(\frac{n_{cr}}{n_{e}}\right)^{1/2}\left(\frac{\lambda}{w_0}\right)^2.
\end{equation}
This results in the magnetic field strength reduction from 0.6 MT to 0.46 MT, which is closer to the value observed in 3D PIC simulations, 0.13 MT. The number of accelerated electrons is also reduced to 34~nC.

Thus, we identified an important parameter governing the effectiveness of the MVA mechanisms, the focusing of the laser. It was already mentioned in Ref. \cite{park2019ion} that in order to maximize the coupling of the laser to the plasma, the laser spot size should be equal to the radius of the self-generated channel. Here, we showed how the magnetic  field, channel radius, and the optimal target thickness are affected when $w_0>R_{ch}$. 

\section{\label{sec:conclusions} Conclusions }

In this paper, we studied the effects of different preplasma scale lengths on the effectiveness of the magnetic vortex acceleration. Previous simulation and analytical studies of this regime of ion acceleration mostly assumed sharp plasma vacuum interface at the back of the target. However, realistic laser temporal contrast conditions can lead to significant target pre-expansion before the arrival of the main laser pulse. Even the best contrast cleaning techniques available now are not able to completely remove the undesirable temporal features. Thus, the interaction with the pedestal and prepulses can lead  to the modification of the target shape and density, resulting in the generation of an expanding preplasma at the front and at the back of the target. We modeled this effect by choosing NCD targets with exponential density ramps of different scale lengths. We performed 3D PIC simulations of an intense laser pulse interaction with these targets and found that the MVA performance depends on the preplasma scale length. While the preplasma at the front of the target might even be beneficial for the laser coupling to the target, as it provides conditions for laser pulse self-focusing and self-steepening, the preplasma at the back of the target has a detrimental effect. We observed that as the scale length was increased, the maximum proton energy decreased. This is linked to the evolution of the accelerating field at the back of the target. For small scale lengths, the accelerating fields are localized near the back of the target and the protons are able to catch up with them and be accelerated. For large scale lengths, the accelerating fields, instead of being localized, continue to move behind the laser pulse and the protons are not able to catch up with them. Thus, we observed the dephasing of protons and accelerating fields, resulting in the reduction of the maximum proton energy. However, in this case, the accelerating and focusing fields are confined within a bubble-like structure while the laser is propagating through the rear surface preplasma, and thus, are enhanced and longer-lived. A suitably structured target, placing or injecting the ions in the correct phase of the fields may considerably enhance the acceleration mechanism.

The accelerated proton beams demonstrate remarkable emittance in the submicron range. The results of 3D PIC simulations for different preplasma scale lengths show the generation of spatially confined, low divergence proton beams, which result in emittance values of $(\epsilon_x,\epsilon_y)\approx(10~\mbox{nm},~20~\mbox{nm})$, where the emittance was calculated for beamlets characterized by a 1~MeV energy spread. Moreover, the emittance evolution during the 0.5~ps time interval as tracked by 3D PIC simulations is quite small, indicating a very limited contribution of the space charge effects.   

The analytical analysis of the MVA, based on the assumptions that the laser pulse is optimally coupled to the target and the number of electrons in the central filament can be determined from the beamloading condition, leads to the following formulae for the peak magnetic field in the channel and the number of accelerated protons. The magnetic field generated by the pinched electron current in the channel scales as 
\begin{equation}
    B_0=0.6 \sqrt{P[PW]}~\text{MT},
\end{equation}
which yields approximately around $0.6$ MT for a 1~PW laser pulse. 
The number of accelerated protons can be estimated as 
\begin{equation}
    N_{max,0}=40(P[PW])^{2/3}(n_e/n_{cr})^{1/6}~\text{nC}.
\end{equation}
This yields approximately 40 nC for a 1 PW laser pulse. The values of $B$ and $N_{max}$ are in good agreement with the results of 3D PIC simulations reported in Ref. \cite{park2019ion}, where optimal coupling is achieved.

We showed, that in the cases studied in this paper, the condition of optimal coupling is not satisfied, i.e., the focal spot size is larger than the radius of the channel that would have been generated without part of the laser being scattered sideways: 
\begin{equation}
    w_0>\frac{2.75}{(n_e/n_{cr})^{1/3}}P[PW]^{1/6} ~\mu m.
\end{equation}
This effect leads to lower values of the magnetic field strength,
\begin{equation}
    B=\frac{2.6}{(n_{e}/n_{cr})^{1/2}w_0[\mu m]^{3/2}}P[PW]^{3/4}~\text{[MT]},
\end{equation}
and the number of accelerated protons,
\begin{equation}
    N_{max}=\frac{300} {(n_{e}/n_{cr})^{1/2}w_0[\mu m]^{2}}P[PW]~\text{[nC]}.
\end{equation}
The ratios of the magnetic fields and numbers of accelerated protons for ideal and non-ideal coupling scale with laser power, plasma density, and laser focal spot size as $B/B_0\sim P^{1/4}n_e^{-1/2}w_0^{-3/2}$ $N_{max}/N_{max,0}\sim P^{1/3}n_e^{-2/3}w_0^{-2}$ respectively. Thus, modification of the laser focal spot size $w_0$ is the most straightforward path to achieving optimal coupling of the laser to the NCD target. 

\section*{Acknowledgments}
This work was supported by the U.S. Department of Energy (DOE) Office of Science, Offices of Fusion Energy Science (FES) and High Energy Physics (HEP), and LaserNetUS, under Contract No. DE-AC02-05CH11231. S. Hakimi was supported by the U.S. DOE FES Postdoctoral Research Program administered by the Oak Ridge Institute for Science and Education (ORISE) for the DOE. ORISE is managed by Oak Ridge Associated Universities (ORAU) under Contract No. DE-SC0014664. This research is based upon work supported by the Defense Advanced Research Projects Agency via Northrop Grumman Corporation. WarpX was supported by the Exascale Computing Project (17-SC-20-SC), a collaborative effort of two U.S. DOE organizations (Office of Science and the National Nuclear Security Administration). This research used resources of the Oak Ridge Leadership Computing Facility at the Oak Ridge National Laboratory, which is supported by the Office of Science of the U.S. Department of Energy under Contract No. DE-AC05-00OR22725.
An award of computer time was provided by the ASCR Leadership Computing Challenge (ALCC) program.
This research used resources of the National Energy Research Scientific Computing Center (NERSC), a Department of Energy Office of Science User Facility using NERSC award FES-ERCAP0027627. All opinions expressed in this paper are the author's and do not necessarily reflect the policies and views of DOE, ORAU, or ORISE.

\bibliography{ref}

\end{document}